\documentclass[prb,twocolumn, showpacs,superscriptaddress, preprintnumbers,amsmath,amssymb]{revtex4}

\usepackage{graphicx,color}
\usepackage{multirow}

\begin{document}

\title{Magnetoresistence engineering and singlet/triplet switching in \\ InAs nanowire quantum dots with ferromagnetic sidegates}

\author{G. F\'{a}bi\'{a}n}
\affiliation{
Institute of Physics, University of Basel, Klingelbergstrasse 82, 4056 Basel, Switzerland\\}
\author{P. Makk}
\affiliation{
Institute of Physics, University of Basel, Klingelbergstrasse 82, 4056 Basel, Switzerland\\}
\author{M.H. Madsen}
\altaffiliation[present address: ]{Danish Fundamental Metrology, DK-2800 Kgs. Lyngby, Denmark}
\affiliation{Center for Quantum Devices and Nano-Science Center, Niels Bohr Institute, University of Copenhagen, Universitetsparken 5, DK-2100
Copenhagen, Denmark}
\author{J.~Nyg{\aa}rd}
\affiliation{Center for Quantum Devices and Nano-Science Center, Niels Bohr Institute, University of Copenhagen, Universitetsparken 5, DK-2100
Copenhagen, Denmark} 
\author{C. Sch\"onenberger}
\affiliation{
Institute of Physics, University of Basel, Klingelbergstrasse 82, 4056 Basel, Switzerland\\}
\author{A. Baumgartner}
\email{andreas.baumgartner@unibas.ch}
\affiliation{
Institute of Physics, University of Basel, Klingelbergstrasse 82, 4056 Basel, Switzerland\\}
\date{\today}

\begin{abstract}
We present magnetoresistance (MR) experiments on an InAs nanowire quantum dot device with two ferromagnetic sidegates (FSGs) in a split-gate geometry. The wire segment can be electrically tuned to a single dot or to a double dot regime using the FSGs and a backgate. In both regimes we find a strong MR and a sharp MR switching of up to 25\%  at the field at which the magnetizations of the FSGs are inverted by the external field. The sign and amplitude of the MR and the MR switching can both be tuned electrically by the FSGs. In a double dot regime close to pinch-off we find {\it two} sharp transitions in the conductance, reminiscent of tunneling MR (TMR) between two ferromagnetic contacts, with one transition near zero and one at the FSG switching fields. These surprisingly rich characteristics we explain in several simple resonant tunneling models. For example, the TMR-like MR can be understood as a stray-field controlled transition between singlet and a triplet double dot states. Such local magnetic fields are the key elements in various proposals to engineer novel states of matter and may be used for testing electron spin-based Bell inequalities.
\end{abstract}

\pacs{85.75.-d, 75.47.-m, 75.76.+j, 73.23.-b}

\maketitle

\subsection{Introduction}
Magnetoresistence (MR) is the electrical resistance of a material or a device as a function of an external magnetic field $B$. For non-magnetic bulk materials one typically finds a parabolic field dependence and a saturation, but it can also be linear to large fields, for example due to disorder,\cite{Kozlova_Eaves_Patane_NatureComm_2012} or a linear dispersion relation,\cite{Abrikosov_PRB58_1998} for example on the surface of a topological insulator.\cite{Zhang_PRL106_2011} Many MR effects -- and applications -- are known for magnetic materials,\cite{Fert_RMP80_2008, Zutic_DasSarma_RevModPhys76_2004} for example the anisotropic magnetoresistance (AMR) used to  characterize magnetic contacts,\cite{Aurich_Baumgartner_APL97_2010, Samm_Gramich_Baumgartner_JAP_2014} or the tunneling magnetoresistance (TMR), in which electrons tunnel between two magnetic reservoirs, resulting in a resistance that depends on the relative orientation of the reservoir magnetizations. TMR typically shows hysteretic rectangular MR loops in up and down sweeps of $B$. Since the domain structure of a magnet can change abruptly with the external field, a characteristic MR switching (MRS) occurs at the respective switching fields $B_{\rm sw}$.

\begin{figure}[b]
	\centering
		\includegraphics{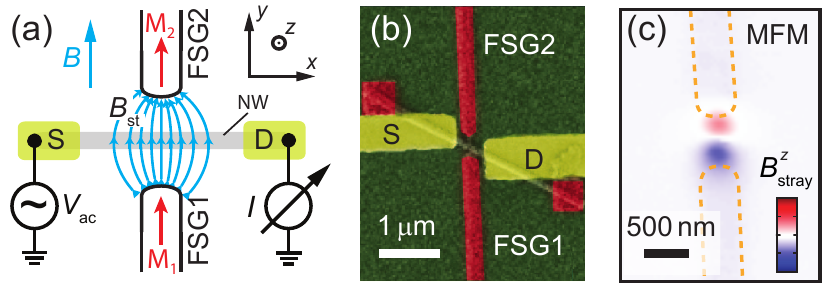}
	\caption{(Color online) (a) Schematic of a NW with ferromagnetic sidegates (FSG1 and FSG2) that result in a local stray field $B_{\rm st}$. The magnetizations $\vec{M}$ of the FSGs can be inverted by an external field $B$. (b) False-color scanning electron microscopy image of the actual device. (c) Magnetic force microscopy (MFM) image showing the magnetic field component perpendicular to the wafer surface of two Py strips outlined by dashed lines and oriented as the FSGs in (b).}
\end{figure}

In nanostructures another type of low-field MR results from the Zeeman splitting of the energy dependent density of states (DOS). For example, we show below how the transport resonances of a quantum dot (QD) can be used as a sensor for the local magnetic field. In turn, this allows one to tune the QD spectrum using the stray-fields of micro- and nanomagnets, without the interface issues associated with magnetic tunnel contacts. Prominent examples are to selectively address quantum bits using stray field gradients,\cite{Pioro-Ladriere_Tarucha_NaturePhys_2008} or spin resonance experiments in double QDs (DQDs).\cite{Forster_Ludwig_PRB91_2015} Nanomagnets are also in the focus of recent proposals to generate helical magnetic fields equivalent to a synthetic spin-orbit interaction that supports, for example, topological quasi-particles like Majorana Fermions in semiconducting nanowires (NWs).\cite{Klinovaja_Loss_PRL109_2012}

Here we report the MR of a NW segment between a ferromagnetic split-gate for different NW transport characteristics, from a single QD to DQDs. The device idea is illustrated in Fig.~1a. The split-gate consists of a pair of long, narrow ferromagnetic Permalloy (Py) strips which we call ferromagnetic sidegates (FSGs). The strips have a single domain magnetization with a switching field determined by the geometric width. \cite{Aurich_Baumgartner_APL97_2010, Samm_Gramich_Baumgartner_JAP_2014} In addition, these FSGs can be used individually as electrical gates. We stress that the NW is not in contact with the FSGs, so that the transport through the NW  is  affected only by the FSG stray field.
Large-diameter NWs can be tuned by electrical gating into different regimes, probably aided by potential fluctuations on the substrate, similar as in two-dimensional electron gases with a continuous transisiton from a single QD to a DQD.\cite{Livermore_Westervelt_Science_1996} We investigate the MR of a single QD (R1), two QDs coupled in series (R2) and two more strongly side-coupled QDs (R3) on the same NW segment. The latter is essentially a parallel DQD in which only one dot is connected to source and drain, which is possible in finite-width NWs and the asymmetric gating of a backgate. We find that the sign and amplitude of the MR and the MRS at $B_{\rm sw}$ of the FSGs can be tuned by the electrical gates, with switching amplitudes of up to $\sim25\%$. In regime R3, we even find a rectangular MRS of $\sim50\%$, reminiscent of TMR, but considerably larger than expected for Py-based TMR devices. For each regime we interpret the MR and MRS using an intuitive resonant tunneling model.

\subsection{Device fabrication}
Figure~1b shows a false-color scanning electron microscopy (SEM) image of the device we discuss here. It consists of a $\sim80\,$nm diameter InAs NW deposited by spin-coating on a Si$^{++}$/SiO$_{2}$ substrate and contacted by source (S) and drain (D) Ti/Au (5~nm/70~nm) contacts. The NWs were grown by solid-source molecular beam epitaxy,\cite{Madsen_Nygard_JCrystGrowth_2013} implementing a two-step growth process to suppress stacking faults.\cite{Shtrikman_Heiblum_Nanolett_2009} We use standard lock-in techniques to measure the differential conductance $G=dI/dV$ as a function of the bias $V$ at a base temperature of $230\,$mK in a $^3$He cryostat. If not stated otherwise, the backgate voltage and dc bias are set to $0$ for all presented experiments, while the electronic structure of the NW segment is electrically tuned by the FSGs. Two $\sim 200\,$nm wide Py strips with a large aspect ratio\cite{Samm_Gramich_Baumgartner_JAP_2014} are fabricated with the ends in close proximity to the NW segment between S and D. The strips are aligned in parallel to the external magnetic field, which results in the angle to the randomly oriented NW axis. The magnetic force microscopy image of an identical split-gate geometry in Fig.~1c demonstrates that the FSG stray field (out of plane component) is confined to a very small volume in the split-gate gap. Two metal pads at the NW ends are used to pin down the NW, which allows a better fabrication accuracy. 

\subsection{Single dot MR and MRS}
We first discuss the MR in the single QD regime (R1). Figure~2a shows $G$ as a function of the FSG voltages $V_{\rm SG1}$ and $V_{\rm SG2}$. We find well-defined Coulomb blockade (CB) resonances with a typical broadening of $\Gamma\approx 200\,\mu$eV. In the gate voltage interval R1 the CB resonances depend continuously on both gate voltages with similar lever arms and a constant slope, characteristic for a single QD formed between S and D. In contrast, for roughly $V_{\rm FSG1}<-0.3\,$V, $G$ exhibits strong avoided-crossings, typical for DQDs.
\begin{figure}[t]
	\centering
		\includegraphics{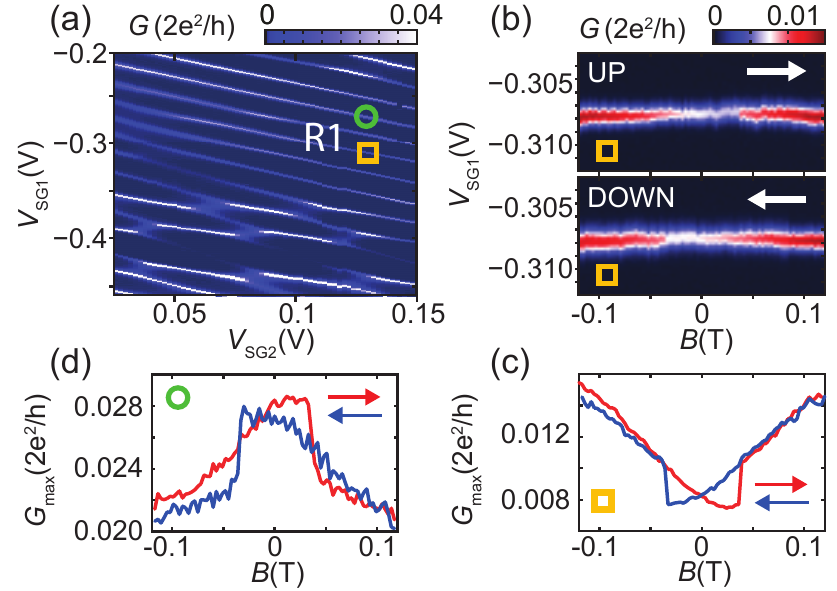}
	\caption{(Color online) (a) Differential conductance $G$ as a function of the voltages $V_{\rm SG1}$ and $V_{\rm SG2}$ applied to the two FSGs. This plot exhibits two regions with characteristics of a single QD (top part) and a DQD (bottom part). (b) $G$ as a function of $V_{\rm SG1}$ and the external magnetic field $B$ for the gate voltages pointed out by a yellow square in (a). The top and the bottom panel show an up and a down sweep, respectively. (c) Maximum conductance $G_{\rm max}$ vs. $B$ for the resonance in (b). (d) $G_{\rm max}$ vs. $B$ for the resonance labeled by a green circle in (a).}
\end{figure}
In Fig.~2b, $G$ of a single QD CB resonance in regime R1 is plotted versus $V_{\rm SG1}$ and the external magnetic field $B$. The measured resonance is pointed out by a yellow square in Fig.~2a. The CB resonance amplitude extracted from Lorentzian fits to vertical cross sections is plotted in Fig.~2c as a function of $B$. When the external field is swept from negative to positive values (UP), we find a gradual reduction of the peak conductance and a sharp switching to a larger value at $B\approx 35\,$mT. In the down sweep (DOWN) $G$ is reduced gradually down to $B\approx-35\,$mT, where again a sharp switching to a larger value occurs. The MRS in this case is $\Delta G / G_{1} \approx +20 \%$, where $\Delta G>0$ is the change in $G$ at the MRS in sweep direction and $G_1$ the conductance just before the switching. We note that the energy of the resonance does not change significantly at these low fields within the measurement accuracy, and that there is a hysteretic increase in the resonance broadening for larger field values. In addition, the sign and amplitude of the MR and MRS depend on the QD resonance. For example, a similar experiment on a different resonance (pointed out by a green circle in Fig.~2a) shows a continuous decrease in conductance for increasing fields and a sharp drop in conductance at the switching fields ($\Delta G <0$), as shown in Fig.~2d. This corresponds to a negative MRS of $\Delta G / G_1 \approx -20 \%$. We can thus tune the sign of both, the MR and the MRS, by choosing the corresponding QD resonance, and on a single resonance by changing one gate voltage and compensating with the other (not shown). However, in regime R1 we could not find a clear systematic dependence on the gate voltages, on which we further comment below.

All experiments presented here can be understood based on the MR of the NW segment in the different regimes. The magnetic field at the NW position is a superposition of the external field and the FSG stray field. The switching field is the same in all experiments on this device and coincides with the FSG magnetization inversion expected from the designed Py strips.\cite{Aurich_Baumgartner_APL97_2010, Samm_Gramich_Baumgartner_JAP_2014} Since the external field is aligned with the FSGs, we assume that in average the stray field points in the same direction and thus adds an offset field, $B_{\rm tot}=B+B_{\rm st}$. For a given dependence $G(B)$ one thus obtains a switching at $B_{\rm sw}$, which shifts the original $G(B)$ curve by the stray field $\pm B_{st}$, with the sign determined by the sweep direction. More details on the switching field are discussed in the next section.

Two basic mechanisms can lead to MR in the single QD regime: 1) the wave function and thus the tunnel couplings $\Gamma_1$ and $\Gamma_2$ to the NW segment can change due to a perpendicular momentum component (Lorentz force)\cite{Vdovin_Patane_Eaves_Science290_2000} or the formation of Landau levels, which in turn changes the maximum conductance, with an MR and MRS sign depending on the asymmetry of the tunnel barrier strengths. These mechanisms, however, typically become relevant at much higher fields. 2) In electronic structures with an energy dependent transmission, the Zeeman shift of the resonances renders the conductance field dependent. Here, we focus solely on the latter possibility and use the simple picture of resonant tunneling through a QD with two spin channels. The conductance then reads
\begin{equation}
	G(B,V_{\rm g})=\frac{e^2}{h}\frac{\Gamma_1 \Gamma_2}{\Gamma}\sum_{\sigma\in\{+,-\}}\mathcal{L}(E_0+\tfrac{1}{2}\sigma g^*\mu_{\rm B}B),
\end{equation}
with $\sigma=+$ ($-$) for spin $\downarrow$ ($\uparrow$), the Lorentzian $\mathcal{L}(E)=\frac{\Gamma}{E^2+0.25\Gamma^2}$, $\Gamma=\Gamma_1 + \Gamma_2$ and the individual spin-independent QD tunnel coupling strengths $\Gamma_1$ and $\Gamma_2$. The argument of $\mathcal{L}$ is the level position, determined by the level energy at zero field $E_0=-e\alpha V_{\rm g}$, with $\alpha$ the gate lever arm and $g^*$ the effective $g$-factor.

In this model the conductance amplitude has a maximum when the two spin states are degenerate and is reduced when this degeneracy is lifted by the Zeeman shift. For a single level the degeneracy occurs generally at $B=0$. At higher fields, the maximum drops to half the value when the orbital energies are fully separated in energy. This would result in a maximum MR and MRS of $\Delta G / G=50\%$ on a field scale $\Delta B=\Gamma/(g^{*}\mu_{\rm B})$. The presented experiments do not reach such large MR values at the relatively low fields, probably because of the large broadening $\Gamma$. We note that this interpretation is not altered in the constant interaction model for a QD with a finite charging energy, though the CB resonance separation is much larger than the Zeeman splitting. A large QD charging energy can be included using a simple two-channel rate equation model for a single QD level, for which we obtain a maximum MR (and MRS) of $25\%$ in the limit of $\Gamma_1=\Gamma_2$, and between $0$ and $50\%$ for strongly asymmetric tunnel barriers. The negative MR and MRS in the single dot regime (e.g. Fig.~2d) cannot be explained by a single level in this model. However, we also find CB conductance maxima at finite fields, which we interpret as level crossings with other orbital states, on the field scale $\Delta B \sim \delta E/(g^{*}\mu_{\rm B})$, with $\delta E$ the level spacing. The observed minima roughly agree with the level spacing found in Coulomb diamond measurements  and an average $g$-factor of $\sim 10$.\cite{dHollosy_Baumgartner_AIPProc_2012}

\subsection{Double dot MR and MRS}
 
\begin{figure}[b]
	\centering
		\includegraphics{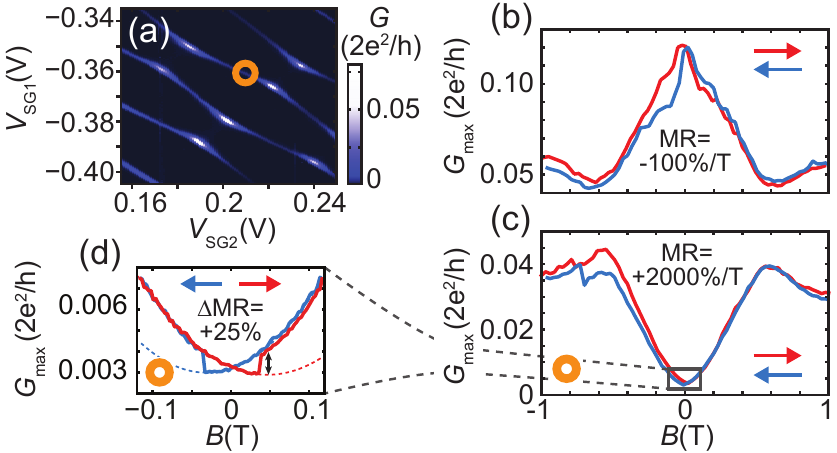}
	\caption{(Color online) (a) $G$ as a function of $V_{\rm SG1}$ and $V_{\rm SG2}$ in the DQD regime R2. (b) and (c) large-field MR on and off a DQD triple point, respectively. The latter gate configuration is pointed out by a yellow circle in (a). (d) low-field MR of the resonance in (c), showing the MRS off a triple point.}
\end{figure}
The same NW segment can be tuned by more negative FSG voltages to a regime with a conductance pattern characteristic for two QDs in series (regime R2).\cite{Wiel_RevModPhys75_2003} In Fig.~3a, $G$ shows a honeycomb pattern with avoided crossings and a $\sim 10$-fold conductance increase near the triple points. In this regime we find very reproducible MR and MRS, with characteristics determined by the gate voltage settings relative to a triple point. We first investigate the MR on a larger field scale. The CB resonance amplitudes on a triple point are generally suppressed with increasing magnetic field. An example is shown in Fig.~3b, where the CB amplitude shows a strong maximum at $B=0$ and is reduced essentially linearly by more than a factor 2, which corresponds to an MR of roughly $-100\%/$T. The local minimum around $B\approx\pm 0.6\,$T is then followed by another maximum at $B\approx\pm1.0\,$T.

In contrast, for MR curves off a triple point we consistently find curves that are all very similar to the one shown in Fig.~3c for the gate voltage configuration indicated by a yellow circle in Fig.~3a. Here, $G$ shows a strong minimum at $B=0$ and an initially parabolic increase by more than a factor of $10$ up to maxima at $B\approx\pm 0.6\,$T, the same field as where the local minima in Fig.~3b occurred. The MR slope on this field scale is roughly $+2000\%/$T relative to the zero-field value. On a smaller field scale we find in all curves an MRS consistent with the high-field MR and a switching of the stray field due to the magnetization reversal in the FSGs. For the example in Fig.~3c, a low-field MR curve is plotted in Fig.~3d. A parabolic fit to the low-field side for the up-sweep (red) and to the high-field side for the down-sweep (blue) is added as a guide to the eye. In an up-sweep, the stray field of the FSGs is oriented in the negative direction for low and negative fields and the resulting total field would be compensated only at a positive external field, $B=B_{\rm st}$. This results in a shift of the parabola by $B_{\rm st}$ (red parabola). For this shift we find in all experiments $B_{\rm st}\approx 50\,$mT. At the switching field $B_{\rm sw}$ the magnetizations in the FSGs reverse and the stray field points along the external field, so that the conductance curve `jumps' to the parabola shifted by $-B_{\rm st}$ (blue parabola). For the switching field we obtain $B_{\rm sw}\approx 35\,$mT, in accordance with the experiments above and previous measurements on non-interrupted long Py strips.\cite{Aurich_Baumgartner_APL97_2010, Samm_Gramich_Baumgartner_JAP_2014} These curves clearly show the difference between the switching and the stray field: the apex of the parabolas are shifted by the stray field given by the geometry and the magnetization, while the switching field is determined by the shape anisotropy of the FSGs.
%
\begin{figure}[t]
	\centering
		\includegraphics{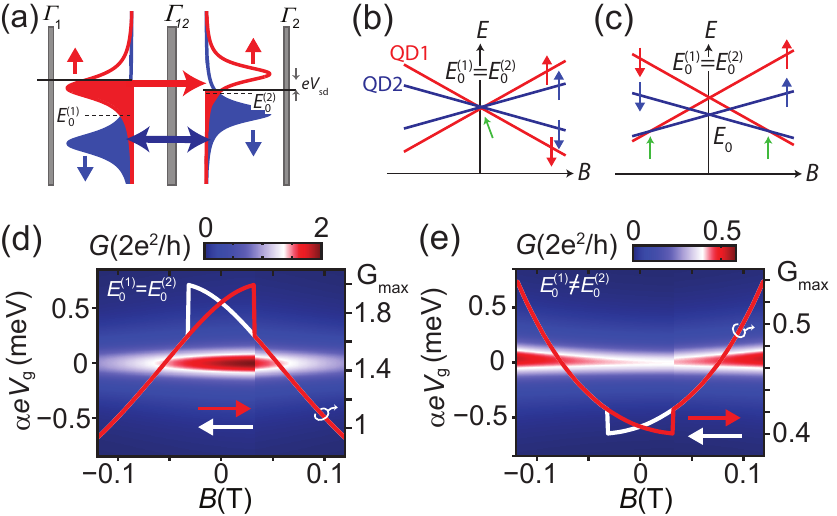}
	\caption{(Color online) (a) Spin-split energy levels in the left and right QD. (b) and (c) Illustrations of the QD level energies for (b) aligned and (c) non-aligned zero-field QD states. The green arrows point out the magnetic fields of resonant tunneling through the DQD. (d) and (e) Calculated conductance $G$ as a function of the $B$ and the gate voltage for resonances (d) on and (e) off a triple point. The overlayed curves show the maximum conductance $G_{\rm max}$ vs. $B$.}
\end{figure}

A qualitative understanding of the DQD MR can be gained by considering the limit of weakly coupled dots. A weak inter dot tunnel coupling compared to the tunnel rates to the leads, $\Gamma_{12}\ll\Gamma_1,\Gamma_2$, suggests that the transmission though the DQD is limited by $\Gamma_{12}$ and the individual QDs can be assumed to be either filled or empty. This picture is illustrated in Fig.~4a. If in addition the capacitive coupling between the QDs is weaker than to the leads and gates, then each QD level, i.e., $E_{0}^{(1)}$ and $E_{0}^{(2)}$, can be tuned individually, for example by local gates.\cite{Fulop_dHollosy_Baumgartner_PRB90_2014} The conductance of the system can then be approximated by
%
\begin{equation}
\begin{gathered}
G(B)= \frac{e^2}{2\pi h^3}\Gamma_1 \Gamma_2 \Gamma_{12}\times \\
\sum_{\sigma\in\{+,-\}} \mathcal{L}(E_0^{(1)}+\tfrac{1}{2}\sigma g^{*}\mu_{\rm B}B)\mathcal{L}(E_0^{(2)}+\tfrac{1}{2}\sigma g^{*}\mu_{\rm B}B)
\end{gathered}
\end{equation}
A triple point in the $B=0$ stability diagram corresponds to both QD levels being aligned with the Fermi energy of the contacts, i.e. $E_{0}^{(1)}=E_{0}^{(2)}=E_{\rm F}$. A finite field separates the two spin states as illustrated in Fig.~4b, and the resonance amplitudes decrease. If both QD states have the same $g$-factor, the Zeeman shift is the same and one expects a decrease only by a factor of 2 on the field scale $\Gamma/(g^{*}\mu_{\rm B})$. However, the two QDs can also have different $g$-factors, either due to anisotropies in the $g$-tensor,\cite{dHollosy_Baumgartner_AIPProc_2012} or due to specific QD wave functions.\cite{Csonka_Hofstetter_NanoLett8_2008} In this case the Zeeman splitting leads to a misalignment of the two QD resonances, which strongly suppresses the transmission with increasing $B$, corresponding to $100\%$ MR. This additional suppression we expect on the larger field scale of $2\Gamma / (|g_{1}^{*}-g_{2}^{*}|\mu_{\rm B})$.

If the two QD levels, $E_{0}^{(1)}$ and $E_{0}^{(2)}$, are different at $B=0$, i.e., if the resonances at zero field are detuned from any triple point, there are characteristic fields where the current is maximized when the Zeeman shift aligns two equal spin levels of the QDs, as pointed out by the green arrows in Fig.~4c, on a field scale governed by $2\Delta E_{0} / (|g_{1}^{*}-g_{2}^{*}|\mu_{\rm B})$ and $\Gamma$. Figures.~4d and 4e show the resulting DQD conductance on and off a zero-field triple point, respectively. This model reproduces qualitatively the characteristics found in the experiments, including an idealized sharp magnetization switching and large MR and MRS values.
We note that in both discussed cases this model predicts strong spin polarized currents, in principle up to $100\%$ at the fields pointed out in Fig.~4c.

\subsection{FSG induced singlet to triplet switching}

At a more negative backgate voltage ($V_{\rm BG}=-0.9\,$V), close to the pich-off voltage, the NW segment exhibits the conductance pattern shown in Fig.~5a (regime R3). The elongated honey-comb structure of the CB resonances are characteristic for a strongly side-coupled parallel DQD,\cite{Andrade_PRB89_2014, Baines_Feinberg_PRB85_2012, Sato_Kobayashi_PRL95_2005} where only one QD is tunnel coupled to source and drain and transport occurs solely at gate voltages where the CB resonances of this first QD are aligned with $E_{\rm F}$. However, the energy spectrum is still determined by both dots.
%
\begin{figure}[t]
	\centering
		\includegraphics{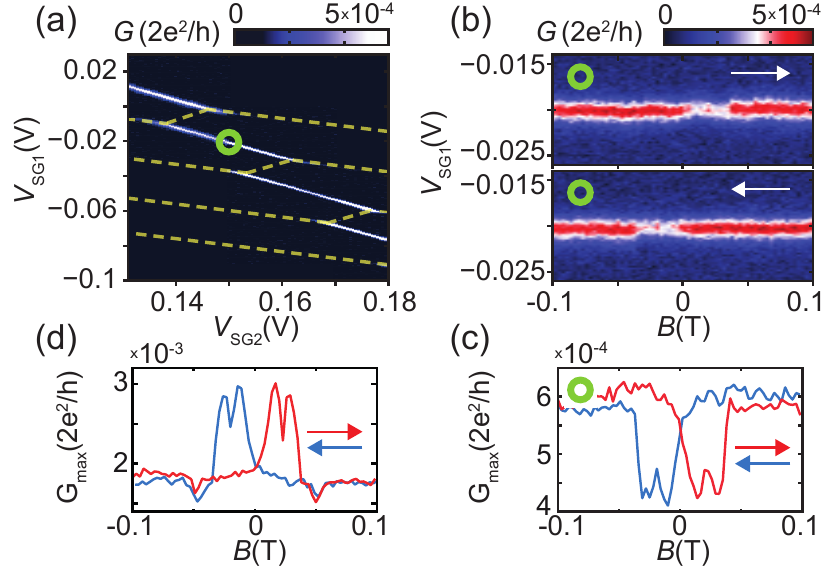}
	\caption{(Color online) (a) $G$ as a function of $V_{\rm SG1}$ and $V_{\rm SG2}$ in regime R3 at a backgate voltage $V_{\rm bg}=-0.9\,$V. The dashed lines highlight the DQD charging diagram. (b) $G$ as a function of $B$ and $V_{\rm SG1}$ in an up- (upper panel) and a down sweep (lower panel) at the gate voltages indicated in (a) by a green circle. (c) Resonance maxima from (b) vs. $B$. (d) Resonance maximum of a different resonance in regime R3 vs. $B$.}
\end{figure}
For some gate voltages in R3 we find MR and MRS very similar to the ones of the DQD discussed for regime R2 in the previous section, whereas for others no significant MR could be resolved. In addition, we also find a new family of MRS curves that occurs at gate voltages that seem rather random. An example is shown in Fig.~5(b), where $G$ is plotted as a function of $B$ and $V_{\rm SG1}$ for the gate voltages indicated by a green circle in Fig.~5(a). The corresponding resonance amplitude is shown in Fig.~5(c). In contrast to the regimes R1 and R2, there are {\it two} transitions in the conductance amplitude, one around $B=0$ and one at $B\approx\pm 35\,$mT$\,\approx B_{\rm sw}$. The change in conductance is negative and corresponds to an MRS of about $-25$\% at the low-field transition and $+50$\% at the high-field transition (for a more symmetric definition of MRS, $\frac{\Delta G}{G_{\rm low}+G_{\rm high}}$, one obtains the same number for both switchings, namely $\pm 20$\%). For the similarity to TMR signals in spin-valves, we call this rectangular shape TMR-like. We do not find any amplitude or position variation with $B$ on this field scale apart from the transitions in the amplitude. The transition around zero field is smoother than in regimes R1 and R2 and already starts at small negative (positive) fields in the up (down) sweep, while the transitions at the switching fields are significantly sharper.

The sign can also be inverted for such TMR-like MRS, as demonstrated in Fig.~5d for a different gate voltage setting. Here we find an increased conductance in the interval $B\in\left[ 0,\pm 35\,{\rm mT}\right]$, which corresponds to an MRS of roughly $+50$\% at the low-field transition. We note that the MR and MRS are almost identical in the two cases shown in Fig.~5.

This TMR-like MRS is qualitatively different to the MRS in the regimes R1 and R2 since the MR is essentially flat on this low field scale. We now show that these characteristics can be explained by a change of the orbital ground state, from a spin triplet to a singlet and vice versa, induced by the FSGs.
In a DQD with an appreciable inter-dot coupling, the two-electron eigenstates are not simply the product of the individual QD states as assumed in the above model, but three spin triplet states $\left|T_{+}\right\rangle= \left| \uparrow, \uparrow \right\rangle$, $\left|T_{0}\right\rangle= \frac{1}{\sqrt{2}}\left(\left| \uparrow, \downarrow \right\rangle  + \left|\downarrow, \uparrow\right\rangle\right)$ and $\left|T_{-}\right\rangle= \left| \downarrow, \downarrow \right\rangle$, and a singlet state $\left|S\right\rangle= \frac{1}{\sqrt{2}}\left(\left| \uparrow, \downarrow \right\rangle  - \left|\downarrow, \uparrow\right\rangle\right)$. In the resonant tunneling picture we neglect fluctuations to single or multiple QD charge states and focus solely on the magnetic field dispersion of the two-electron states. As illustrated in Fig.~6a, $\left|T_{+}\right\rangle$ and $\left|T_{-}\right\rangle$ are shifted linearly up and down in energy with increasing field, respectively, while $\left|T_{0}\right\rangle$ and $\left|S \right\rangle$ are constant with an energy separation given by the exchange energy $E_{\rm x}$. Without FSG stray fields, all triplet states cross at zero field, and the $\left|T_{\pm}\right\rangle$ cross $\left|S\right\rangle$ at a finite filed, depending on $E_{\rm x}$. One finds that the lowest energy orbital state changes from a triplet to a singlet and back to a triplet at $B=\pm E_{\rm x}/(g^{*}\mu_{\rm B})$.

The effect of a homogeneous FSG stray field is illustrated in Fig.~6b for a down-sweep. The up-sweep can be constructed analogously. While $\left|T_{0}\right\rangle$ and $\left|S\right\rangle$ are unaffected by $B_{\rm st}$, $\left|T_{+}\right\rangle$ and $\left|T_{-}\right\rangle$ are offset in $B$ by $-B_{\rm st}$ for $B>-B_{\rm sw}$ and by $+B_{\rm st}$ for $B<-B_{\rm sw}$. In the example depicted in Fig.~6b, if $B_{\rm st}>B_{\rm sw}$, $\left|T_{+}\right\rangle$ crosses $\left|S\right\rangle$ at $-B_{\rm sw}$ and $\left|T_{-}\right\rangle$ at the characteristic singlet/triplet transition field $B_{\rm tr}=E_{\rm x}/(g^*\mu_{\rm B})-B_{\rm st}$. In this case the ground state is a singlet for $-B_{\rm sw}<B<B_{\rm tr}$ and a triplet otherwise. The second consecutively filled state is a singlet, except for the same interval around zero, and the higher energy states are triplets for all fields.  

%
\begin{figure}[b]
	\centering
		\includegraphics{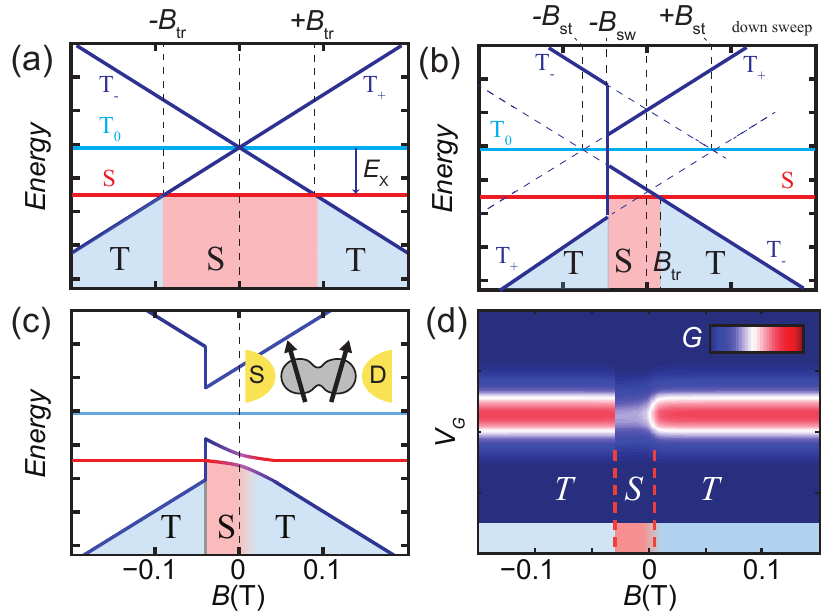}
	\caption{(Color online) (a) Energy levels in a strongly coupled DQD for a homogeneous magnetic field and no FSG stray field. The triplet states are labeled by $T_{+}$, $T_{-}$ and $T_{0}$, respectively, and the singlet by $S$. Where the S and $T_{\pm}$ cross, the lowest energy state changes its character from triplet (T) to a singlet character (S). (b) Similar plot as in (a) with a stray field $B_{\rm st}$ that switches orientation at $B_{\rm sw}$ and a characteristic singlet/triplet transition at $B_{\rm tr}$. (c) Similar as (b) with non-parallel local magnetic fields $B_1$ and $B_2$ on the two QDs, with a $\pm 30^{\circ}$ angle with respect to the external field. (d) Calculated conductance $G$ as a function of $B$ obtained from the DQD model discussed in the text.}
\end{figure}
The singlet/triplet transitions in the picture up to now are very sharp. However, if one accounts for an inhomogeneous stray field of the FSGs - quite expected in this sample because of the tilted orientation of the NW with respect to the FSGs - the eigenstates become coherent superpositions of the singlet and triplet states. The energies of these states depend on the magnetic field at the positions of the two individual QDs, $\mathbf B_1$ and $\mathbf B_2$. The eigenenergies of this system can be found by diagonalizing the Hamiltonian $H$ in the singlet/triplet basis $\{T_{+}, T_{0}, T_{-}, S\}$ \cite{Petersen_Ludwig_PRL110_2013}
\begin{equation}
H=g^{*}\mu_{\rm B} \left(
\begin{array}{ccccc}
	\overline{B}_{z} & \frac{1}{\sqrt{2}}\overline{B}_{-}  & 0								 & -\frac{1}{\sqrt{2}}\Delta B_{-} \\
	\frac{1}{\sqrt{2}}\overline{B}_{+} & 0 								 & \frac{1}{\sqrt{2}}\overline{B}_{-} & \Delta B_{z} \\
	0								& \frac{1}{\sqrt{2}}\overline{B}_{+} & -\overline{B}_{z}	 & \Delta B_{+} \\
	-\frac{1}{\sqrt{2}}\Delta B_{+} & \Delta B_{z} & \Delta B_{-} & E_{\rm x}/g^{*}\mu_{\rm B}
\end{array}
\right),
\end{equation}
with $\overline{{\mathbf B}}=\frac{1}{2}\left( {\mathbf B}_{1}+{\mathbf B}_{2}\right)$ the average and $\Delta{\mathbf B}=\frac{1}{2}\left( {\mathbf B}_{1}-{\mathbf B}_{2}\right)$ the difference of the fields at the two QD positions, and $\overline{B}_{\pm}=\overline{B}_{x}\pm\overline{B}_{y}$ and $\Delta B_{\pm}=\Delta B_{x}\pm\Delta B_{y}$, with the $z$-axis defined along the external field. $E_{\rm x}$ is the exchange energy.
In Fig.~6(c) we plot the eigenenergies as a function of the external magnetic field in a down-sweep for $E_{\rm x}<0$. We add a constant stray field of the strength found in the experiments, with the respective angles $\pm 30^{\circ}$ to the external field at the two QD positions. At $B_{\rm sw}$, we reverse the orientation of $B_{\rm st}$. The non-parallel fields at the QD positions result in an anti-crossing of the $\left|T_{-}\right\rangle$ and $\left|S\right\rangle$, as can be seen in Fig.~6(c). With lowering $B$ the weight of the singlet increases steadily around $B_{\rm tr}$. However, at $-B_{\rm sw}$ the ground state is switched back to $\left|T_{+}\right\rangle$ sharply due to the magnetization reversal, which makes the system "jump over the anti-crossing" one would expect at negative fields.

This singlet/triplet transition can be observed in a conductance measurement because the respective orbital wave functions have different spatial symmetry, so that one can expect different tunnel couplings and thus different conductances for these states. A switching of the ground state from a triplet to a singlet (and back) thus results in a switching of the device resistance. As an illustration, we plot in Fig.~6d the calculated conductance for the energies in Fig.~6c, with a Lorentzian gate dependence and the maximum conductance $G_{\rm max}=\left|\eta_{\rm S}\right|^2 G_{\rm S}+\left|\eta_{\rm T}\right|^2 G_{\rm T}$, where $\gamma_{\rm S,T}$ are the the superposition amplitudes obtained from diagonalizing $H$. $G_{\rm S}\neq G_{\rm T}$ are the characteristic conductances of the singlets and triplets states, respectively. This simple model accounts qualitatively for the TMR-like MRS, with a sharp singlet/triplet switching at $-B_{\rm sw}$ and a smoother transition around $B=B_{\rm tr}\approx 0$. In this picture, the high-field ground state of the next charge state is the singlet, which switches to a triplet for the same field interval given above, which accounts for a negative TMR-like MRS. The next two charge states would then be expected to show no MR or MRS, in accord with measurements at other gate voltages.

The seemingly random characteristics with gate voltage might to some part originate from an uncertainty of the relative positions in the charge stability diagram. However, the TMR-like MRS also requires a fine-tuning of the device parameters, such that $B_{\rm st}>B_{\rm sw}>B_{\rm tr}$, which can be inferred directly from Fig.~6b. Here the stray field $B_{\rm st}$ is given by the FSG spacing and material choice, the switching field $B_{\rm sw}$ by the FSG shape anisotropy (strip width), and the singlet/triplet transition field $B_{\rm tr}$ by the exchange energy, i.e. the DQD coupling strength. The seemingly random occurrence of the TMR-like MR might therefore also originate from state-to-state fluctuations in $B_{\rm tr}$, possibly related to NW $g$-factor fluctuations and anisotropy.\cite{Csonka_Hofstetter_NanoLett8_2008, dHollosy_Baumgartner_AIPProc_2012} In addition, different  spatial probability distributions of different QD wave functions would lead to different relative angles of the local magnetic fields $B_1$ and $B_2$ and to strong variations in the observed MRS patterns.


\subsection{Summary and outlook}
In summary, we report magnetoconductance measurements of a short segment of InAs nanowire with a pair of ferromagnetic side gates in a split-gate geometry. For this device we find three characteristic regimes, which we identify as a single QD, a double dot in series and a strongly side-coupled DQD. While in the former two a strong field dependence of the conductance results in a simple single MRS due to the inversion of the FSG magnetization, we find in regime R3 a double switching reminiscent of a TMR signal. The MR and MRS in all three regimes we explain qualitatively in simple resonant tunneling models. Especially in regime R3 we can reproduce qualitatively the data by assuming molecular DQD states with transitions between orbital states of singlet and triplet character, induced by the FSG stray fields.
As an outlook we first note the versatility of FSG structures: FSGs of different widths should result in different switching fields and in switchable and strong field gradients. Second, we note that in our experiments we have not reached the limits of how close FSGs can be fabricated to a NW and we have not explicitly exploited the angle between the NW and the FSGs, which might be used to tailor a specific field modulation along the NW. A closer proximity should result in considerably larger stray fields, on the order of several hundred mT. This and the straight forward scalability to long FSG arrays are promising for creating synthetic spin-orbit interactions and helical electron states. In addition, we propose to use FSGs as the basis for an all-electronic Bell test, for example in a NW-based Cooper pair splitter.\cite{Hofstetter2009, Hofstetter_Baumgartner_PRL107_2011, Das_Heiblum_NatureComm_2012, Fulop_dHollosy_Baumgartner_PRB90_2014} Similar as the spin-orbit interaction in Ref.~\citenum{Braunecker_Levi_Yeyati_PRL111_2013}, one might use the large $g$-factor in NWs and the FSG stray fields $B_{\rm st}$ to obtain a Zeeman splitting larger than the thermal energy and the life time broadening of a quantum dot (QD), $g^{*}\mu_{\rm B}B_{\rm st}>k_{\rm B}T, \Gamma$, leading to spin-polarized QD states. The local magnetic fields can then be designed to results in different orientations of the quantization axes on the two QDs, determined by the vectorial sum of $B_{\rm st}$ and the external field. This setup avoids fundamental problems with electronic Bell tests based on ferromagnetic contacts. \cite{Klobus_Baumgartner_Martinek_PRB89_2014}

\subsection{Acknowledgements}
We thank J. Schwenk and H.-J. Hug from EMPA for the MFM image and S. d'Hollosy and S. Zihlmann for helpful discussions. This work was financially supported by the EU FP7 Project SE2ND, the EU ERC
Project QUEST, the Swiss National Science Foundation (SNF), including
the projects NCCR QSIT, and the Danish Research Councils.

\bibliographystyle{apsrev}

\end{document}